\documentclass[11pt]{article}
\usepackage{amssymb,amsmath,amscd,latexsym}
\begin{document}


\title{REPLY TO COMMENTS OF BASSI, GHIRARDI, AND TUMULKA ON THE FREE WILL THEOREM}
\author{John Conway$^\star$ and Simon Kochen\thanks{jhorcon@yahoo.com; \
$^\star$kochen@math.princeton.edu}\\
Princeton University, Department of Mathematics\\ Princeton, NJ
08544-1000}

\maketitle

\newcommand{\ra}{\mathrm A}
\newcommand{\rb}{\mathrm B}
\newcommand{\iap}{\mathrm A'}
\newcommand{\ibp}{\mathrm B'}

\parskip .15cm

\begin{flushleft}
{\bf 1.  Introduction}       
\end{flushleft}

We reply to the recent comments on our paper [1] by Bassi and
Ghirardi [2] and Tumulka [3], and thank them for helping us explain
our ideas more clearly. We briefly summarize their criticisms
and our responses to them.

They both start by equating our FIN axiom with Bell's locality
condition; we agree that the latter is false, but argue that the former
is true, since it follows from relativity and causality.

Bassi and Ghirardi also questioned whether their "hits" should count as 
information subject to FIN - we recast our proof below to make it clear 
that all that matters is that the responses of either particle cannot 
vary with the choices of the other's experimenter (this is the new 
assumption, MIN, that replaces FIN, causality,and the Free Will 
assumption). It appears that Bassi and Ghirardi do not accept even this, 
although it follows immediately from relativistic invariance and 
experimental free will.

Some of Tumulka's other criticisms involve the fact that the functions
controlling his flashes might be frame dependent.  This does no harm to
our argument, because it used the particle responses, which cannot be
frame dependent even if the flashes are.  Indeed frame dependence,
rather than removing contradictions, produces even more!

We now assent to Tumulka's argument that the question of adding an 
interaction term is irrelevant. However we show that, like the types of 
theory that Bassi and Ghirardi propose, rGRWf must fail MIN, despite 
Tumulka's assertion to the contrary.

\begin{flushleft}
{\bf 2. Recasting the proof}
\end{flushleft}

We first recast our proof to make it clear that concerns about
inertial frame dependence and exactly what counts as information are
irrelevant.

The postulated functions
\begin{displaymath}
   \theta_a^{\Lambda_1}(x,y,z,w,\alpha) \qquad \text{and} \qquad
\theta_b^{\Lambda_2}(x,y,z,w,\beta)
\end{displaymath}
are now allowed to depend on possibly different inertial frames 
$\Lambda_1$ and $\Lambda_2$, and $\alpha$ and $\beta$ are information 
available to the two particles before their responses (in the respective 
frames). Information that becomes available to $a$ only after the choice 
of $x,y,z$ is still treated as at the end of the proof in [1], and so is
not considered here.

We have also introduced $x,y,z,w$ as new arguments to $\theta_a$ and
$\theta_b$ (whether they vary with them or not). Then, as in Section 3
of [1], we clear $\alpha$ and $\beta$ of any dependence on $x,y,z,w$ by
replacing any information-bit $i$ that depends on $x,y,z,w$, by the
values $i_1, ... , i_{1320}$ it takes for the $40 \times 33 = 1320$
particular instances of these variables we use. 
This yields a further renaming
of the $\theta_a^{\Lambda_1}(x,y,z,w,\alpha')$ and
$\theta_b^{\Lambda_2}(x,y,z,w,\beta')$
in which $\alpha'$ and $\beta'$ are now independent of $x,y,z,w$.

\begin{flushleft}
{\bf 3.  Our minimal assumption}
\end{flushleft}

A careful examination of [1] reveals that relativity, causality, FIN,
and the Free Will Assumption were used only to derive the following
``minimal assumption," which now completely replaces them.

{\it The MIN axiom}.  $B$ can freely choose any one of the $33$
particular directions $w$
independently of $\beta'$. Also, neither this choice nor $b$'s response
can be affected by $A$'s choice of $x,y,z$. Similarly $A$ can choose any
one of the $40$ triples $x,y,z$ independently of $\alpha'$, and neither
this choice nor $a$'s response can be affected by $B$'s choice of $w$.

The reason is twofold. It is $B$'s free will that allows him to choose
$w$ independently of the earlier information $\beta'$. But in a suitable
frame $A$'s experiment will happen only five minutes later that $B$'s,
and so $w$ and $b$'s response must also be independent of $A$'s choice
and $a$'s response (for otherwise causality would be
violated).\footnote{We remind the reader that in [1] causality was explicitly assumed.  Without this assumption, what we show is that a relativistically invariant GRW theory has the defect that no matter how we associate inertial frames to particles, some particle's behavior at a certain time in its frame must be affected by events that occur only at a later time.}

The recast statement is that the axioms SPIN, TWIN and MIN already
imply that the responses of the particles cannot be given by functions 
of the information accessible to them (i.e., that is earlier in some 
arbitrary pair $\Lambda_1, \Lambda_2$ of inertial frames).\footnote{We remark that if, as in [1], we require $\theta_a$ and $\theta_b$ to depend only on information in backward light cones, then correspondingly we need only require the free decisions to be independent of information in those cones.}

\begin{flushleft}
{\bf 4. Finishing the proof}
\end{flushleft}

The assumption MIN ensures that the values of $\theta_a$ and
$\theta_b$ (being the responses of the two particles) are in fact
independent of  $w$  and $x,y,z$ respectively, so we can simplify their
names by omitting those arguments:
\begin{displaymath}
  \theta_a^{\Lambda_1}(x,y,z,\alpha') \qquad \text{and} \qquad
\theta_b^{\Lambda_2}(w,\beta')\,.
\end{displaymath}
Now there is a value $\beta_0$ of $\beta'$ for which
$\theta_0(w) = \theta_b^{\Lambda_2}(w,\beta_0)$
is defined for one choice of $w$, and so, by MIN, for all 33. Similarly
there is a value $\alpha_0$ of $\alpha'$ for which 
$\theta_a^{\Lambda_1}(x,y,z,\alpha(0))$
is defined for one of the triples $x,y,z$, and so by MIN for all 40.

Finally, by TWIN, we have (using the question-mark convention of [1])
\begin{displaymath}
  \theta_a^{\Lambda_1}(x?,y,z,\alpha_0) = \theta_0(x),\;
  \theta_a^{\Lambda_1}(x,y?,z,\alpha_0) = \theta_0(y),\;
  \theta_a^{\Lambda_1}(x,y,z?,\alpha_0) = \theta_0(z),\;
\end{displaymath}
(these responses being independent of frame), and by SPIN,
$\theta_0(w)$ is a 101-function, which contradicts the Lemma in
Section 2 of [1]. This completes the proof.

\begin{flushleft}
{\bf 5.  FIN Versus Bell locality}
\end{flushleft}

Our critics both start by equating FIN with Bell's locality condition.
We were aware of course that Nature has non-local correlations - indeed
TWIN expresses one such, so that Bell locality is false. But we argued
that FIN is true, since it follows from causality and relativity, and so
it is wrong to equate FIN with Bell locality.
By doing so, our critics have failed to appreciate that
compared to Bell's theorem, ours has extra strength 
which is needed for the application to GRW.

However, since FIN has now been weakened to MIN, that discussion is now
irrelevant. This also makes it clear that Bassi and Ghirardi's concerns
about which information is subject to FIN are immaterial - our new
condition MIN is just that $a$ may not use $w$ nor $b$ use $x,y,z$.

It seems that Bassi and Ghirardi do not accept even this:
\begin{quote}
``$\ldots$ the outcome of the measurement which $A$ performs on particle
$a$ {\em does} -- indeed it {\em must} -- depend on the outcome of the
measurement performed by $B$ on $b$ (or vice-versa), in particular on
the choice of directions made by $B$ $\ldots$"
\end{quote}

In the context of their discussion of ``relativistic GRW models" this
admission is quite astonishing. For let us suppose that $a$'s response
is conditioned by B's decision. Then in a frame in which $B$'s
experiment happens only 5 minutes later than $A$'s, this can be regarded
either as a gross violation of causality or a restriction on $B$'s
freedom (since he may make only those choices that are compatible with
the already given response).

\begin{flushleft}
{\bf 6.  Tumulka's further criticisms}
\end{flushleft}

The above was also the first of Tumulka's three supposed flaws in our
proof.  His ``third flaw" distinguishes between saying in every frame, or
in just one frame, that the information $\alpha'$ that determines $a$'s
response is independent of $w$.  The new proof obtains a contradiction
from any one pair of frames for which the particles' responses are
determined by functions $\theta_a^{\Lambda_1}$ and $\theta_b^{\Lambda_2}$.

The essential point is that $a$ has only one response, which is
independent of frame. The contradiction is multiplied, rather than
removed, by introducing frame dependence.

Tumulka's ``second flaw" is similar. He says that ``every frame
$\Lambda$ provides a different choice of the function $f_y^{\Lambda}$
[that determines the flash $f_y$]."  But even if this is so, the
particle responses that the flashes supposedly determine cannot vary
with the frame. Tumulka finds it acceptable for his flashes to ``entail
influences to the past in [some] frames," but we cannot accept that
future experimental decisions can influence the past responses of
particles as exhibited by macroscopic spots on screens.

Tumulka allows the function $f_y^{\Lambda}(F_A, F_B, X_1, X_2,...)$ that
gives the flash controlling one of the two particles to depend on both
experimenters' choices $F_A$ and $F_B$. This would seem to make his
theory rGRWf fail MIN, but in fact he claims that there is one frame
which the flash $f_A$ does not depend on the field $F_B$ and another in
which $f_B$ and does not depend on $F_A$ (see his Sections 8 and 7). If
true, then MIN would hold in rGRWf, but then our recast proof applies
and produces a contradictory 101-function.

\begin{flushleft}
{\bf 7.  Invariance}
\end{flushleft}

Unlike Tumulka, who expects his final theory to be fully Lorentz
invariant, Bassi and Ghirardi admit that their proposed type of theory
``will not be invariant in the ordinary sense," but only in a weaker,
stochastic, sense.  We welcome this frank confession
that indeed a GRW theory cannot be fully Lorentz invariant, which was
all that we claimed.  What can this weaker, stochastic, sense of
invariance mean?  Not that the described kind of theory is itself
Lorentz invariant in {\em any} sense, as is shown by their statement
that ``the jumps propagate instantaneously".  We presume that they are
merely claiming that its {\em predictions} will be (stochastically)
invariant.

They justify this by claiming that this type of theory is as
invariant as QM is.  This is simply wrong; the predictions of QM are, to
a high degree of approximation, fully invariant, not just stochastically
so, and in [1], we took care to state the ones we used in an invariant
manner: that if and when measurements of both particles in a given
direction have both been made, the responses will 
coincide.\footnote{Our insistence upon the relativistically correct way of viewing the EPR experiment is hardly new.  Every careful textbook on QM stresses the fact that observables of a system take values only on decoherent interaction with another system such as a measuring apparatus. We quote from the book [4] in which Bohm (before he became a Bohmian!) introduced the spin version of EPR: ``Thus, for a given atom, {\em no} component of the spin of a given variable exists with a precisely defined value until interaction with a suitable system, such as a measuring apparatus, has taken place. $[\dots]$ Thus, in every instance in which particle No. 1 develops a definite spin component in, for example, the $z$-direction, the wave function of particle No. 2 will automatically take such a form that it guarantees the development of the opposite value of $\sigma_z$ if this particle also interacts with an apparatus which measures the same component of the spin.''} 
A credible explanation of how reduction actually takes place should
have the property that not only its predictions, but the theory itself
should be as invariant as Nature seems to be.

References

[1] J. Conway, S. Kochen, Found. Phys. 36 (10), 1441 (2006).

[2] A. Bassi, G. C. Ghirardi, arXiv: quant-ph/0610209 (1 Dec 2006).

[3] R. Tumulka, arXiv: quant-ph/0611283-v2 (6 Dec 2006).

[4] D. Bohm, {\em Quantum Theory} (Prentice-Hall, New York, 1951).

\end{document}